\documentstyle[art10,epsf,wrapfig]{article}

\begin{document}
\def\sc#1{$_{\mbox{\rule{0cm}{0.2cm}#1}}$}
\def\scm#1{_{\mbox{\rule{0cm}{0.2cm}#1}}}
{\small
\vspace{0.2cm}
\centerline {\Large \bf A New Method to Reconstruct the Energy}
\vspace{0.2cm}
\centerline{\Large \bf and Determine the Composition of Cosmic Rays} 
\vspace{0.2cm}
\centerline {\Large \bf from the Measurement of Cherenkov Light}
\vspace{0.2cm}
\centerline {\Large \bf and Particle Densities 
in Extended Air Showers\footnote[5]{Summary of two talks 
given at the {\it XV${^{th}}$\,Cracow Summer School of 
Cosmology, Lodz, 1996}}}
\vspace*{0.4cm}
\centerline {A.\,Lindner, {\it Uni.\,Hamburg, II.\,Inst.f.Exp.Physik,}}
\centerline{\it Luruper\,Chaussee\,149, D-22761\,Hamburg, Germany,
E-Mail:\,lindner@mail.desy.de}
\begin{abstract}
A Monte-Carlo study is presented using
ground based measurements of the 
electromagnetic part of showers initiated in the atmosphere 
by high energetic cosmic rays 
to reconstruct energy and mass
of primary particles with energies above 
300\,TeV.
With two detector arrays measuring the Cherenkov light and the
particle densities as realized in the HEGRA experiment
the distance to the shower maximum and the lateral development
of air showers can be coarsely inferred.
The measurable shower properties  
are interpreted to determine energy 
and energy per nucleon of the 
primary particle.
\end{abstract}
\section{Introduction}
This paper describes a new method to derive the elemental composition and 
the energy spectrum of cosmic rays
for energies above a few hundred TeV
from measurements of Cherenkov light and particle 
densities at ground level.
The main idea is to determine the distance between the detector 
and the shower 
maximum, which is used to correct experimental observables
for fluctuations in the shower developments.
The depth of the shower maximum itself turns out to be a coarse
measurement of the energy per nucleon of the nucleus hitting the 
atmosphere.
The paper is organized as follows:\\
After describing the experimental observables which will be used to 
reconstruct characteristics of the primary particle,
the event simulation is sketched and features of the longitudinal and 
lateral shower development in the atmosphere are considered.
The following three sections deal with reconstruction of 
the position of the shower maximum, of the energy per nucleon
and of the primary energy.
Finally methods to determine the chemical composition are 
described followed by the summary and conclusions.
\section{The Experiment and  Observables} 
Although the method described in this paper was 
developed primarily for the HEGRA experiment it can equally well be 
applied to any installation registering Cherenkov light and charged 
particles of extended air showers (EAS).
The method can be easily generalized to all experimental setups 
which allow the determination of the distance to the shower maximum.
However some properties of the HEGRA experiment need to be mentioned to 
understand details discussed in the following sections.\\
The experiment HEGRA is a multi-component detector system 
described in detail elsewhere \cite{hegra}
for the measurement of extended air showers (EAS).
At a height of 2200\,m a.s.l.\ it covers an area of 180${\rm \cdot 180\,m^2}$.
In this paper
only the scintillator array of 245 huts with a grid spacing of 15\,m
including denser part near the center and the so called AIROBICC array 
of 72 open photomultipliers measuring the Cherenkov light of 
air showers on a grid with 30\,m spacing 
also with a central concentration, are used.
The energy threshold (demanding a signal from at least 
14 scintillator or 6 AIROBICC huts) 
lies at 20\,TeV for proton and 
80\,TeV for iron induced showers.\\
\label{observables}
The measured particle density in the plane perpendicular to the 
shower axis is fitted by the NKG formula \cite{nkg}.
In the fit a fixed Moliere radius of 112\,m is used.
The shape parameter {\it age} and the integral number of 
particles {\it Ne} result from the fitting procedure.
As the HEGRA scintillators are covered with 5\,mm of lead (which suppresses 
the detection of low energy electrons but 
allows the measurement of photons after pair production in the lead)
the values obtained for {\it age} and {\it Ne} cannot be compared to 
simple expectations 
from the cascade theory:
{\it age} maybe smaller than 1 although the measurement takes place well 
behind the shower maximum 
while the shower size {\it Ne} is generally larger than for measurements 
without lead on the scintillator detectors. \\
The Cherenkov light density is only analyzed in the interval
20\,m ${\rm <}$ r ${\rm <}$ 100\,m from the shower core
due to technical reasons although 
HEGRA in principle could 
sample the Cherenkov light density up to 200\,m. 
In the range between 20 and 100\,m the Cherenkov light density  
can be well described by an 
exponential
\begin{equation} 
{\rm \rho_C(r)={\it a} \cdot exp(r\cdot {\it slope})}.
\end{equation}
As in the NKG fit two parameters are obtained from the analysis 
of the Cherenkov light: 
the shape parameter {\it slope}
and the total number of Cherenkov photons
reaching the detector level between 20 and 100\,m core
distance {\it L(20-100)}.
\section{Simulation}
EAS in the energy range from 300\,TeV to 10\,PeV were simulated using 
the code CORSIKA 4.01 \cite{cors}.
The model parameters of CORSIKA were used with their default values 
and the fragmentation parameter was set to ``complete fragmentation''.
This results in a complete disintegration of the nucleus after 
the first interaction.
Showers induced by the primary proton,
${\rm \alpha}$, oxygen and iron nuclei were calculated.\\
The number of generated Cherenkov photons 
corresponds to a wavelength  
interval of ${\rm [340-550\,nm]}$.
In the main this paper assumes 
perfect measurements 
of the number of particles and Cherenkov photons
and a perfect shower core determination
in order to concentrate on the physical principles and 
limitations of the methods to be described.
To study the influence of the realistic experimental performance
the events were passed through a 
carefully checked detector simulation \cite{volker}
(performed with measured response functions)
and reconstructed 
with the same program as applied to real data.
Here each event was used 20 times to simulate 
different core positions inside and impact points outside
the experimental area, 
which nevertheless fire sufficient huts of the arrays to 
fulfill the trigger conditions.\\
In total 1168 events were generated with CORSIKA\,4.01 
with zenith angles of 0,15,25 and 35$^0$\ at 
discrete energies between 300\,TeV and 10\,PeV.
\section{The Development of Showers in the Atmosphere}
Some basic characteristics of the EAS simulated with 
CORSIKA\,4.01 are summarized here.
Features independent or sensitive to the mass of the 
nucleus hitting the atmosphere are described.
These will allow the reconstruction of  
primary energy and mass from the observables mentioned above. 
\subsection{The longitudinal Shower Development}
\label{long}
Shown on the left of Figure\,\ref{longdev} are 
the mean longitudinal developments 
of 300\,TeV proton and iron induced air showers, where
electrons and positrons above an energy of 3\,MeV were 
counted.
This will be subsequently called, the shape of 
the longitudinal shower development.
For each shower the maximum (defined as the point in the shower
development with the maximal number of particles)
was shifted to zero before
averaging.
Afterwards the mean distribution was normalized to the mean particle number 
at the shower maximum.   
     \begin{figure}[htb]
     \begin{center}
     \vspace{-1.3cm}
     \epsfxsize=10.55cm
     \epsfysize=5cm
     \mbox {\epsffile{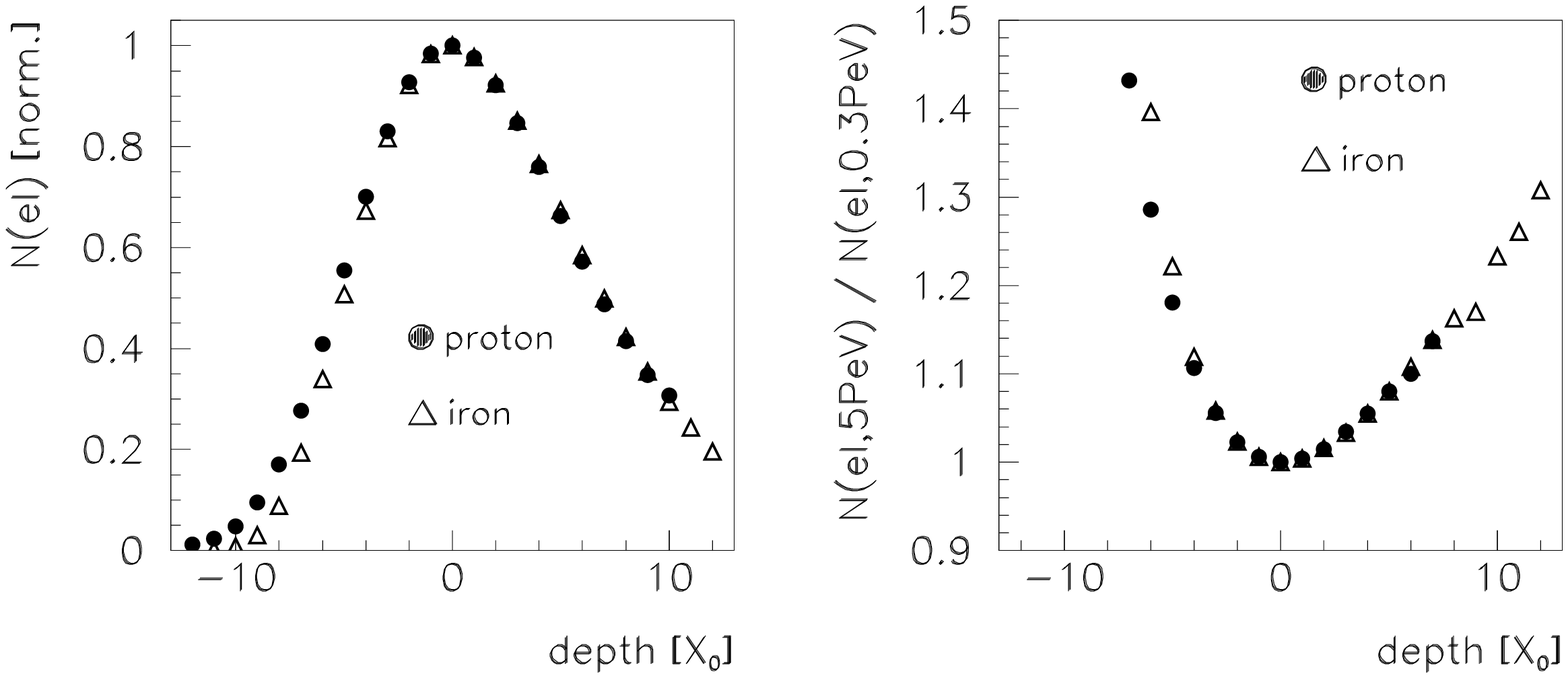}}
     \vspace{-0.5cm}
     \caption{{\small Left: the mean longitudinal 
development of 300\,TeV p and Fe 
showers normalized to the number of particles in the maximum.
The depth of each shower maximum was shifted to 
0\,${\rm X_0}$.
Right: the mean longitudinal development for 
5\,PeV showers divided by the mean 
development for 300\,TeV showers (normalized at the 
shower maximum at 0\,X$_0$).}}
     \vspace{-0.5cm}
     \label{longdev}
     \end{center}
   \end{figure} 
Concerning  the shape of the longitudinal
development behind the shower maximum
no systematic differences depending on the primary particle are visible.
The right plot in Figure\,\ref{longdev} shows the change in the 
longitudinal development with increasing primary energy.
The shapes broaden independently of the primary 
particle. 
Both plots may be explained by a lucky combination of two effects: 
\begin{enumerate}
\item As visible from simulated proton showers at 300\,TeV and 5\,PeV
the longitudinal shower shape gets broader with increasing energy. 
\item After the first interaction an iron induced shower 
can be described as a superposition of
nucleon induced subshowers.
Each of them have different subshower maxima fluctuating around a mean value.
To achieve the distributions in
Figure\,\ref{longdev} the maxima of the whole EAS (and not the maxima of the subshowers)
were overlayed.
Therefore a Fe shower
appears to be broader than 
a proton shower of the same energy per nucleon. \\
\end{enumerate}
In CORSIKA 4.01 (other simulations have to be tested)
both effects combine in such a way that the longitudinal
shape of the EAS behind the maximum
becomes independent of the mass of the primary nucleon for the same
primary energy.\\
The mean atmospheric depths of the maxima depend on the energy per nucleon 
E/A (see section\,\ref{detea}) and are subjected to large fluctuations.
Figure\,\ref{detenuc} shows the corresponding correlation:
the column density traversed by a shower up to its maximum, 
named depth of maximum in the following
(calculable from the distance and the zenith angle), is correlated 
with E/A for all different simulated primaries from p to Fe 
and all zenith angles.
     \begin{figure}[htb]
     \begin{center}
     \vspace{-1.3cm}
     \epsfxsize=10.55cm
     \epsfysize=5cm
     \mbox {\epsffile{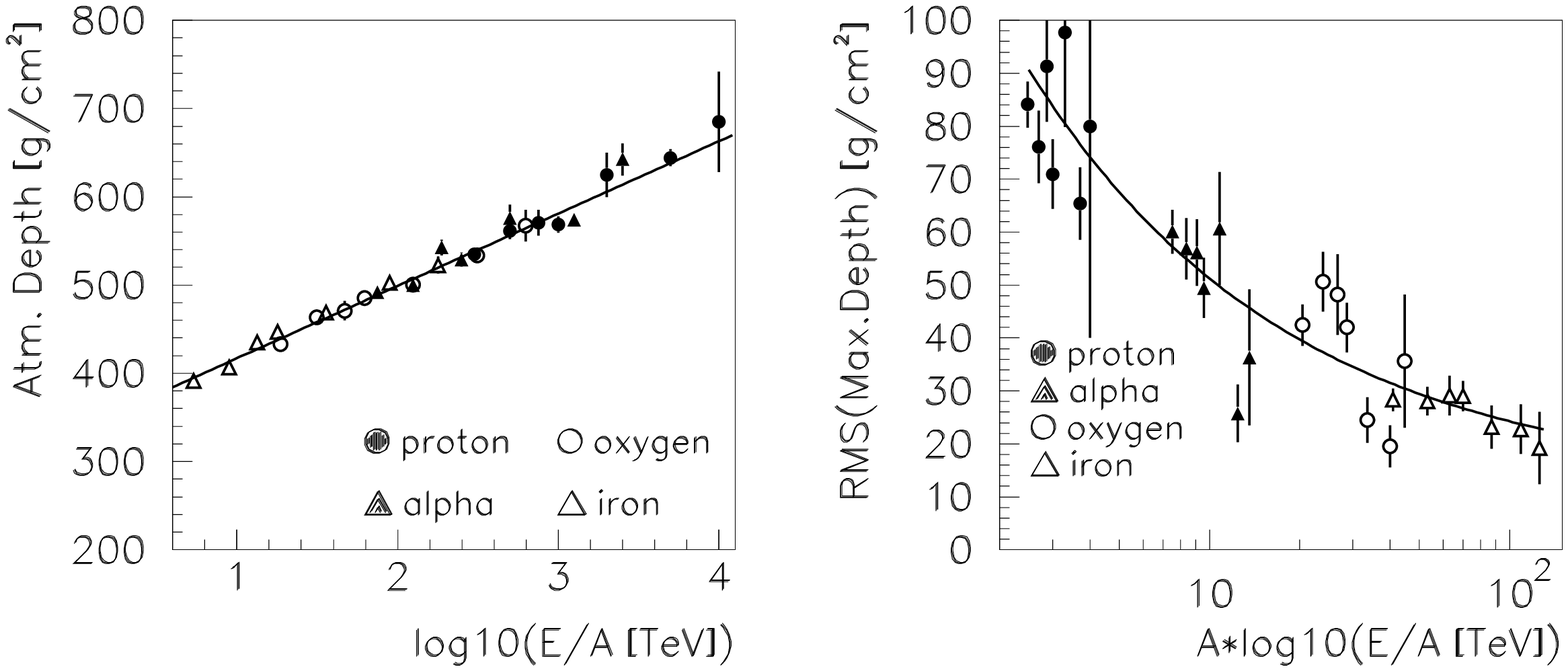}}
     \vspace{-0.5cm}
     \caption{{\small Left: the mean atmospheric depth of the 
shower maxima as a function of energy per nucleon.
The line shows a fit to the correlation.
Right: as an attempt to illustrate the origin of  
the fluctuations (rms) of the  atmospheric depth of the 
shower maxima they are plotted as a function of 
energy E and nucleon number A.
The line shows a fit to the correlation.}}
     \vspace{-0.5cm}
     \label{detenuc}
     \end{center}
   \end{figure} 
With the ``complete fragmentation'' option in our simulations
the correlation follows a linear function.
From Figure\,\ref{detenuc} an elongation rate of approximately 
82\,g/cm$^2$/log$_{10}$(E/E$_0$) is derived: 
\begin{equation}
{\rm depth(max)\,=\,\left[ (335 \pm 3) \,+\,(82 \pm 2) \cdot log_{10} 
\left(\frac{E/A}{TeV}\right)\right]\,g/cm^2 .}
\label{eqenuc}
\end{equation}
If the depth of the shower maximum is measured the 
energy per nucleon E/A can be inferred.
Due to statistical fluctuations in the shower development
and therefore in the depth of the shower maximum
(Figure\,\ref{detenuc})
the resolution for E/A is modest.
The spread decreases with increasing nucleon number A
as the EAS of a complex nucleus consists of 
many overlapping nucleon induced subshowers so that the 
whole EAS exhibits less fluctuations 
than the individual subshowers.    
The resolution improves slightly also with 
rising E/A because more interactions take place 
until the shower maximum is reached.
Figure\,\ref{detenuc} (right) shows a 
parameterization of the fluctuations of the depth of 
the shower maxima.\\
It is interesting to note that for a specific primary particle 
and energy the number of particles in 
the shower maximum Ne(max) is independent of the depth of the maximum.
Therefore this number differs between proton and iron induced showers
of the same primary energy 
even if the shower maxima are accidently at the same position. \\
The most important characteristics of the longitudinal
shower development discussed above and which will be used
for the reconstruction
of primary energy and mass in the next sections are: 
\begin{itemize}
\item The longitudinal shower development behind the shower 
maximum does not depend on the mass of the primary particle
and only slightly on the primary energy.
\item The mean depth of the shower maximum is determined only by
the energy per nucleon E/A.
\item Fluctuations in the position of the shower maximum 
decrease with increasing nucleon number 
and slightly with increasing energy.
\end{itemize}
\subsection{The lateral Shower Development}
\label{showerlat}
In hadronic interactions the typical transverse momentum 
stays roughly constant with energy.
Therefore the lateral spread of 
hadronic showers should decrease with increasing energy 
per nucleon as 
the ratio of transverse to longitudinal momentum gets smaller
in the early part of the shower development
where the energies of the interacting particles are still 
comparable to the primary energy.
In principle this effect could be measured 
for a known distance to the shower maximum
i.e.\ by comparing 
the number of Cherenkov photons reaching the detector
level relatively close to the core with 
the number of all photons 
detectable at the ground level or by analyzing {\it age}.
In this way
a separation of heavy and light primaries turns out to be 
possible at energies below 1\,PeV. 
At energies in the knee region nearly no differences
between proton and iron showers remain.
Obviously here E/A even for iron showers becomes so large that
any influence of the hadronic transverse momentum is washed out 
and the lateral shape of the shower is dominated by 
scattering processes and interactions of particles of 
relatively low energies
in the later part of the shower development.
\section{Reconstruction of the Position
         of the Shower Maximum}
\label{recmax}
This section deals with the reconstruction of the distance between
the shower maximum and the detector.
It can be determined from  
the shape of the lateral Cherenkov light density ({\it slope})
and in principle also from the shape parameter {\it age} of the particle 
distribution at detector level. \\
\label{recmaxcl}
As already noticed by Patterson and Hillas \cite{hillas}
for showers with energies above 1\,PeV
the distance between the detector and the maximum of an EAS 
can be inferred from the lateral distribution of the Cherenkov light
within about 100\,m core distance.
This is possible, because 
Cherenkov light emitted at a specific height shows
a specific lateral distribution at detector level.
The light from the early part of the shower development,
where the energies of the particles are still very high
so that scattering angles are very small,
is concentrated near 120\,m 
(the so called Cherenkov ring). 
Cherenkov light produced closer to the detector level
hits the ground closer to the shower core.
The measurable Cherenkov light density of one EAS
is the sum of all 
contributions from all heights, where lateral distributions from different 
heights enter with amplitudes corresponding to the number of 
Cherenkov light emitting particles in the different heights.
Hence the shape of the measurable lateral light density
distribution depends 
on the longitudinal shower development.
If the shower maximum approaches the observation level  
more light is produced close to the detector reaching the 
ground near the shower core.   
Consequently the lateral Cherenkov light density 
in the range up to 100\,m core distance
drops the steeper the closer the shower maximum approaches the detector.\\
In Figure\,\ref{maxcor} the correlation between the distance to 
the shower maximum and the parameter {\it slope} derived from the 
Cherenkov light distribution is plotted for different primary 
nuclei and zenith angles up to 35 degree
for primary energies of 0.3 and 5\,PeV.    
     \begin{figure}[htb]
     \begin{center}
     \vspace{-1.3cm}
     \epsfxsize=10.55cm
     \epsfysize=5cm
     \mbox {\epsffile{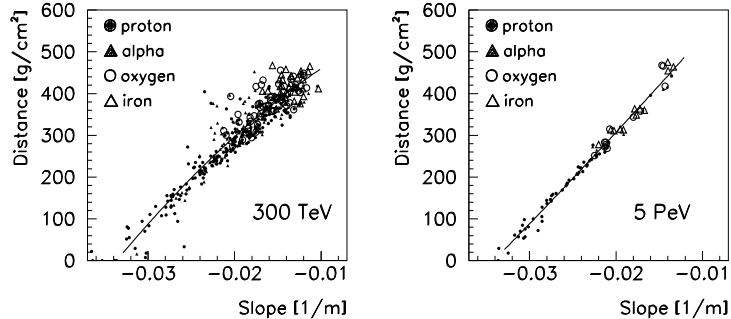}}
     \vspace{-0.5cm}
     \caption{{\small The distance between detector and 
shower maxima plotted against the parameter {\it slope}
for 300\,TeV (left) and 5\,PeV (right) primary energy.
The lines show fits to the correlations.}}
     \vspace{-0.5cm}
     \label{maxcor}
     \end{center}
   \end{figure} 
The distance to the shower maximum can be accurately determined 
from {\it slope}
independent of the type of the primary particle and 
zenith angle\footnote{neglecting atmospheric absorption}
The correlation between distance and {\it slope} depends slightly on the 
primary energy due to the changing 
longitudinal shower development.
Figure\,\ref{maxcor} shows simple polynomial fits describing 
the correlation rather well. 
The dependence of the fit parameters on log(E) were again 
parameterized with polynomials
resulting in a two dimensional function of
{\it slope} and log(E) to determine the distance to the shower 
maximum.
The systematic uncertainties in the reconstruction of the 
distances
depending on particle type or
primary energy are less than 5\% 
increasing a little for zenith angles of 35$^0$.
Such a systematic error for large zenith angles
\label{sysangel}
is expected as these showers develop longer at high 
altitude where the threshold energy for electrons to produce
Cherenkov light is higher than for showers with vertical incidence.\\
Two shower properties contribute to the accuracy of the 
determination of the shower maximum.
For a given primary energy
the resolution improves with decreasing distance between
detector and shower maximum and with increasing
number of nucleons.
      \begin{figure}[tbh]
     \begin{center}
    \vspace{-1.3cm}
     \epsfxsize=10.55cm
     \epsfysize=5cm
     \mbox {\epsffile{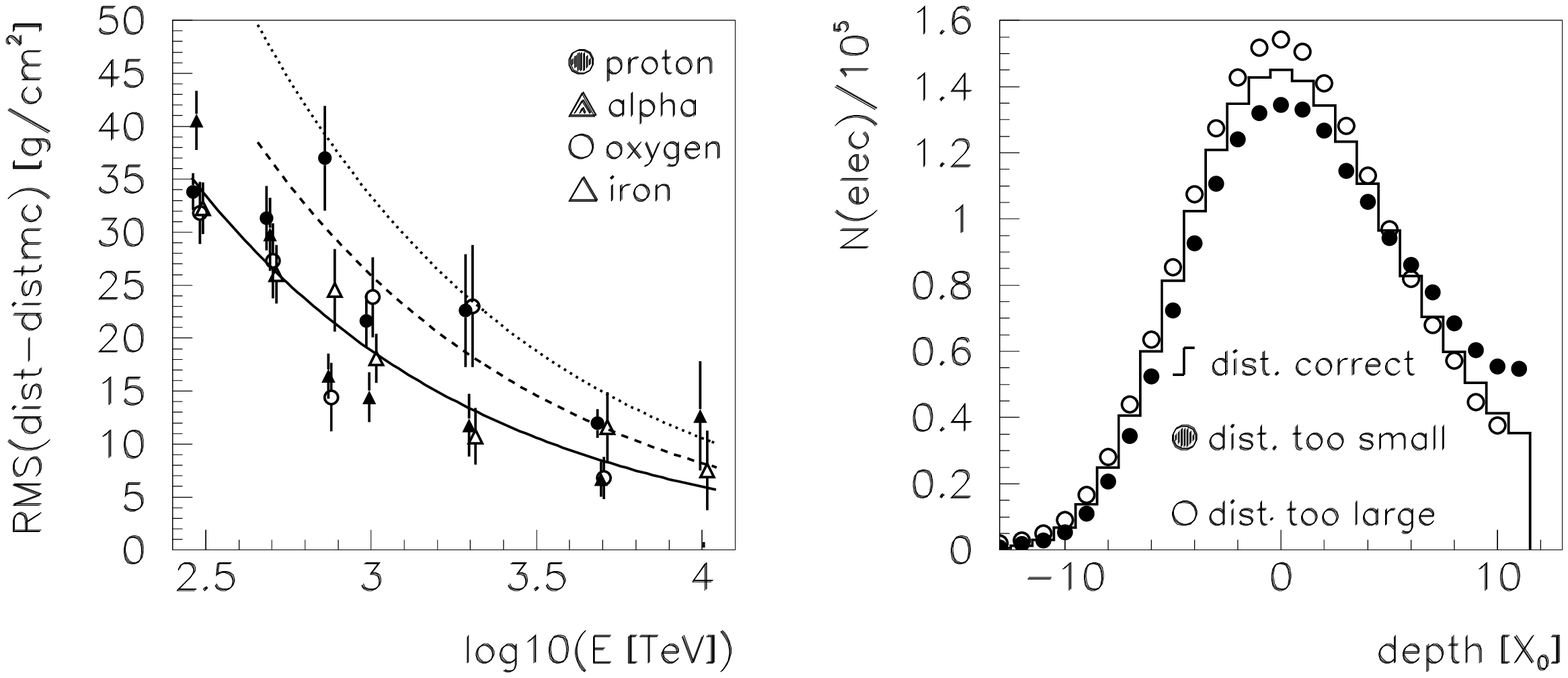}}
     \vspace{-0.5cm}
     \caption{{\small Left: the rms value of the distributions 
of the absolute difference between reconstructed
and MC generated distance to the shower maximum for 
zenith angles of 0$^0$ and 15$^0$. 
The line shows a fit to the correlation.
The broken and dotted lines indicate fits to simulated events
at zenith angles of 25 and 35$^0$\ respectively.\protect\newline 
Right: the mean longitudinal development of 300\,TeV proton 
showers where the reconstructed distance is more than
20\,${\rm g/cm^2}$ too large (open dots), too small (full dots)
or correct within 10\,${\rm g/cm^2}$ (line).
The maximum for each individual shower was shifted to
zero before averaging.
In contrast to Figure\,\ref{longdev} the distributions were not 
normalized to the number of particles in the maximum.}}
     \vspace{-0.5cm}
     \label{distrms}
     \end{center}
   \end{figure} 
Accidentally both contributions behave in such a manner that
for fixed zenith angle and primary energy the resolution becomes 
independent of the mass of the primary particle within the 
statistical errors of the event sample.
Corresponding results are plotted in Figure\,\ref{distrms}.
The right part of Figure\,\ref{distrms} shows the reason for 
the finite resolution of the distance determination with {\it slope}.
Showers, where the distance is underestimated, do not decay as 
fast as an average shower behind the shower maximum.
Consequently {\it slope} is smaller than expected and the distance 
is reconstructed too small.
Showers with overestimated distances exhibit 
a shorter longitudinal extension behind the shower maximum.
It is interesting to note that the length of the shower behind the 
maximum is anti correlated with the number of particles in the 
maximum: the faster the decay after the maximum the more particles 
arise in the maximum. \\
It is worthwhile to note that other approaches,
like the reconstruction of the shower maximum from measurements of the 
time profile of the Cherenkov light pulses
at core distance beyond 150\,m
\cite{volker} show different limitations as 
different shower properties than with {\it slope} are measured.
In principle it is also possible to determine the distance to 
the shower maximum with {\it age},
but the experimental resolution of an {\it age} measurement  
cannot compete with a {\it slope} determination. 
An improvement is only possible with a much denser coverage 
with active detector components than in present experiments
(typically around 1\%).
In addition {\it age} depends on the primary particle type 
so that an unbiased measurement of the position of the shower maximum 
is not possible at energies below 1\,PeV.
\section{Determination of Energy per Nucleon E/A}
\label{detea}
     \vspace{-1.cm}
     \begin{minipage}[t]{6.5cm}
     \vspace*{-8.3cm}
     \epsfxsize=6.5cm
     \epsfysize=4.3cm
     \mbox {\epsffile{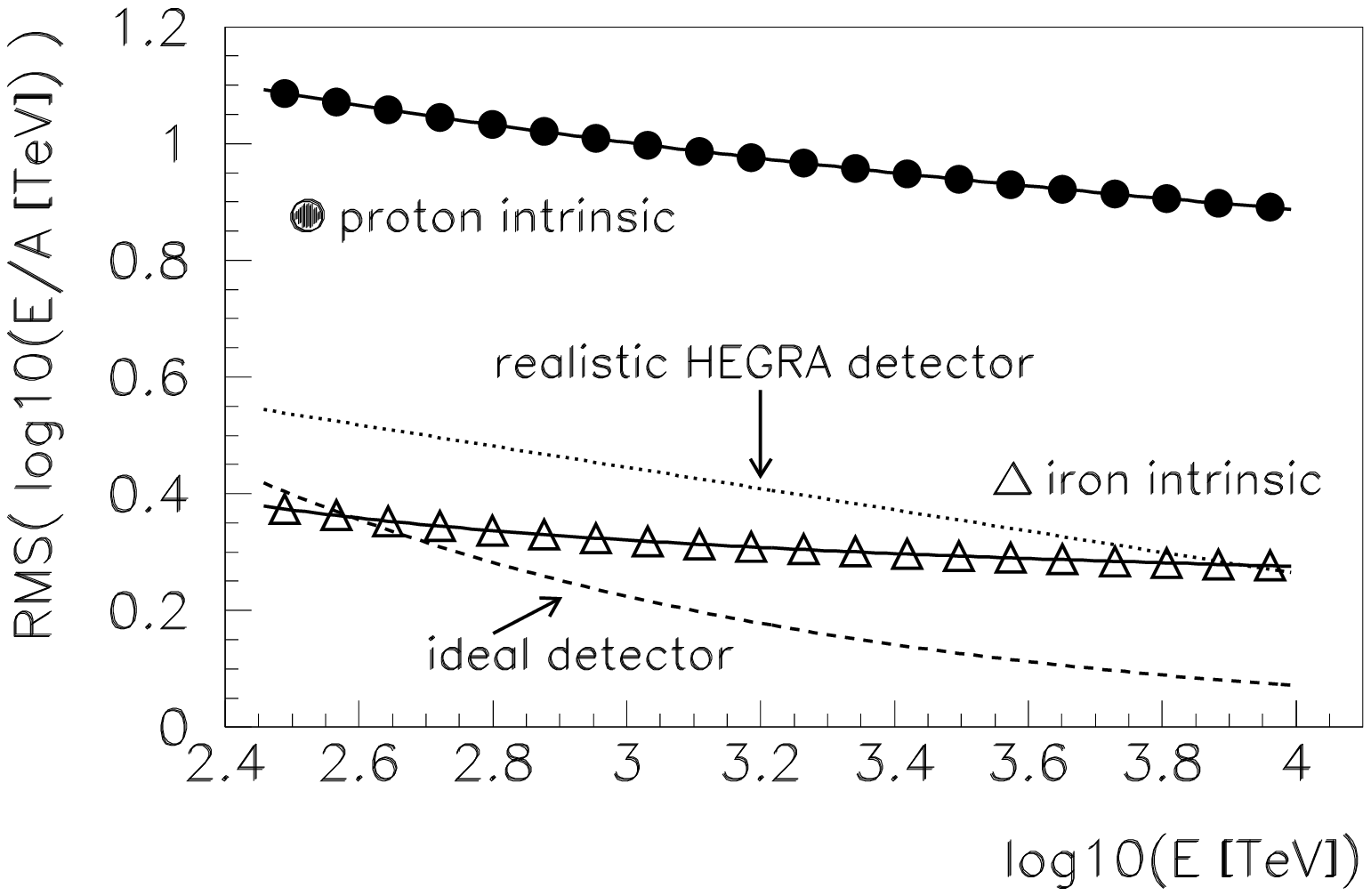}}
{\sl Figure 5: }{\small Different contributions to the accuracy of the 
E/A reconstruction with {\it slope} and zenith angle:
``intrinsic'' contributions refer to Figure\,\ref{detenuc}
showing the fluctuations in the shower development.
The broken 
line ("ideal detector") displays the principal accuracy
limit achievable with the method
to reconstruct the distance to the shower maximum with {\it slope},
the dotted line shows the corresponding accuracy including the 
realistic performance of the present HEGRA
detector
(both for 0 and 15$^0$\ zenith angle).}
     \label{enucrms}
 \end{minipage} \hfill
\begin{minipage}[b]{5.8cm}
\vspace*{1.0cm}
\addtocounter{figure}{1}
With known distance to the shower maximum and known zenith angle
the penetration depth of the shower into the atmosphere until it 
reaches the maximum can be inferred.
E/A can be estimated 
using the correlation shown in Figure\,\ref{detenuc}.
The resolution for E/A is limited by the natural 
fluctuations of the shower maxima positions
(Figure\,\ref{detenuc}) and by principle accuracy limits
of the {\it slope} method (Figure\,\ref{distrms}). 
Of course further experimental errors may 
contribute in addition.
In Figure\,5 the different contributions 
(to be added quadratically)
are compared.
The uncertainty for 
proton induced showers is always dominated by statistical 
fluctuations of the shower developments,
whereas for iron showers at low energies the intrinsic uncertainty 
of the {\it slope} method contributes significantly.
\end{minipage}
\newpage \noindent
At energies around the ``knee'' the accuracy is always limited 
by variations of the shower developments. 
\section{Reconstruction of primary Energy}
With a HEGRA type of detector two different methods may be 
used to reconstruct the primary energy:
\begin{enumerate}
\item Interpreting the shower size at detector level
measured by the scintillator array corresponds to a 
determination of the leakage out of the ``atmospheric calorimeter''.  
These ``tailcatcher'' data allow for an accurate energy 
reconstruction if combined with information on the 
shower development.
This ansatz can be applied as the shape of the longitudinal 
development behind the shower maximum 
does not depend on the primary particle (see 
section\,\ref{long}) and the distance to 
the shower maximum can be measured independently of the
mass of the primary nucleus.
\item The measurement of the amount of Cherenkov light  
makes use of the atmosphere as a fully active calorimeter.
Although in principle superior to the ``tailcatcher'' approach
this idea suffers from the large extension of the Cherenkov light pool.
Depending on the distance to the shower maximum only
20 to 55\% of all Cherenkov photons reach the detector within 
100\,m core distance.
In contrast to this the particle density drops very fast with increasing 
distance to the shower core 
so that a detector of the HEGRA size is sufficient to 
determine the number of all particles reaching the ground level. 
\end{enumerate}
Both algorithms described in the present paper can roughly be divided into 
two steps:
first the distance of the shower maximum to the detector (derived 
from the shape of the lateral Cherenkov light density distribution)
is used to correct for different
longitudinal shower developments.
As only experimental quantities measuring  
the electromagnetic part of the air shower are considered here
it follows naturally, that only the energy 
deposited in the electromagnetic cascade can be reconstructed 
directly.
In a second step a correction for the non measured energy
has to be performed.
This correction depends on E/A only, which is determined from the 
depth of the shower maximum as described in the previous section.\\
The following plots and parameterizations
only take into account showers which reach their maximum at least
50\,g/cm${\rm^2}$ above the detector, because otherwise 
one can hardly decide whether a shower reaches its maximum 
above detector level at all.
The treatment of showers arriving at detector level before reaching
their maximum has to be considered separately.
\subsection{Energy Reconstruction from Particle and \protect\newline
Cherenkov Light Measurement}
\label{ene}
In this section the primary energy will be reconstructed in the 
following manner:
\begin{enumerate}
\item With the known distance to the shower maximum 
the number of particles in the shower maximum Ne(max)
is derived from the measurement at detector level {\it Ne}.
\item Ne(max) is proportional to the energy deposited in the 
electromagnetic cascade.
\item From the energy per nucleon E/A and Ne(max) 
the total primary energy is inferred.
\end{enumerate}
\subsection*{}
\vspace*{-1.cm}
\begin{wrapfigure}[14]{l}[0pt]{7cm}
     \begin{center}
     \vspace{-0.7cm}
     \epsfxsize=6.5cm
     \epsfysize=4.3cm
     \mbox {\epsffile{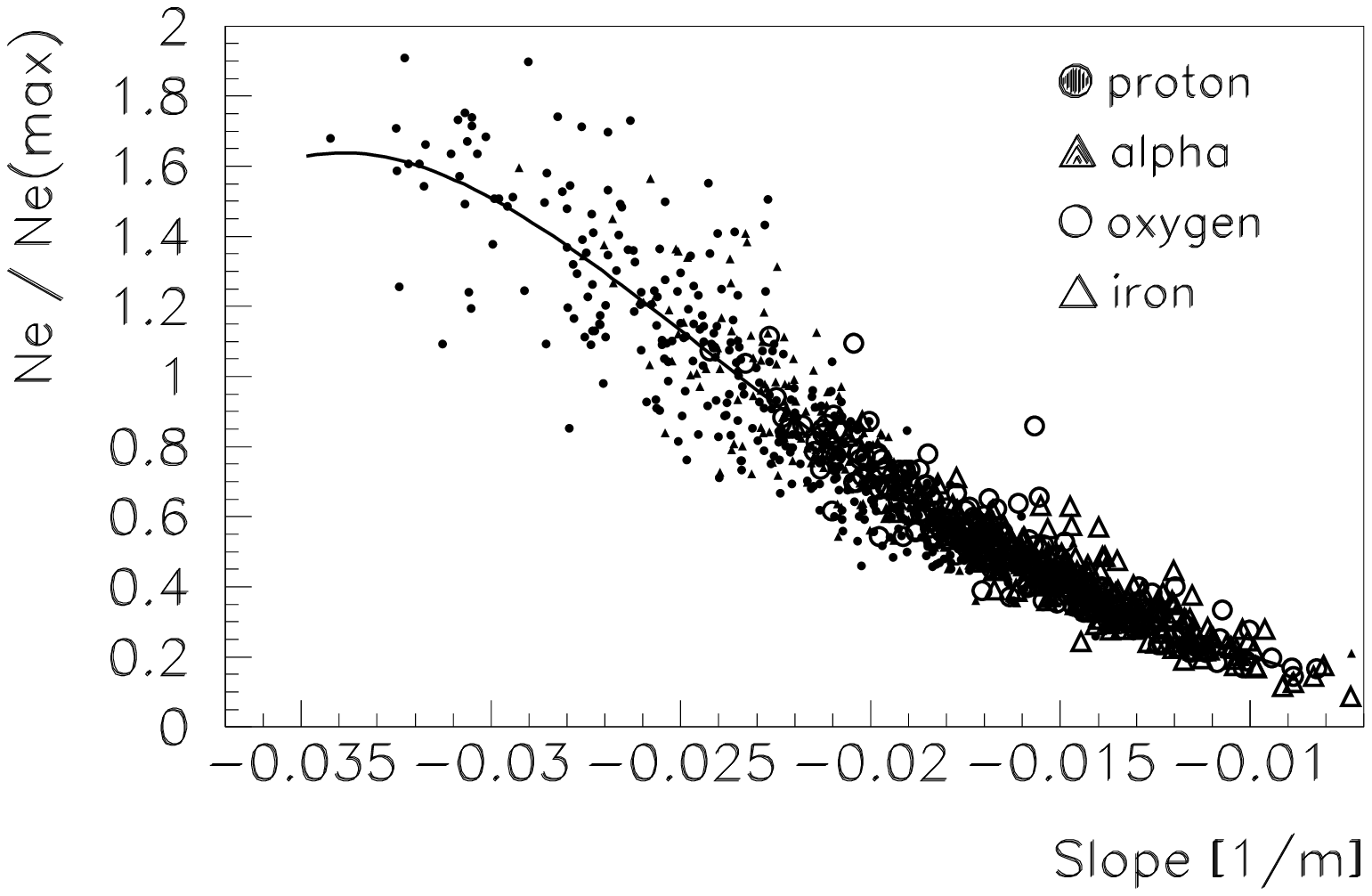}}
     \vspace{-0.5cm}
     \caption{{\small The ratio of {\it Ne} (measured below 0.5\,cm of 
lead) and the 
number of particles at the maximum Ne(max) 
as a function of {\it slope}.
The line shows a fit to the correlation.}}
     \label{eemcor}
     \end{center}
   \end{wrapfigure} 
The first step for reconstructing the primary energy 
is the determination of Ne(max) from the observables
{\it Ne} and {\it slope}.
As the scintillator huts of the HEGRA experiment are 
covered with 1\,${\rm X_0}$ of lead the total number of 
particles is not measured directly. In simulations
the ratio of measured particles and the number of 
particles before the lead was found to depend only on the 
distance to the shower maximum but not on primary energy 
nor on the nucleon number of the primary particle.
Therefore no systematic uncertainties in the energy 
reconstruction originate from the lead coverage.
Due to the conversion of photons in the lead the measured number of particles 
is larger than Ne(max) for showers reaching their 
maximum close to the detector.
In Figure\,\ref{eemcor} the ratio of {\it Ne} to 
{\rm Ne(max)} 
is correlated with {\it slope} measuring the 
distance between detector and shower maximum.
Using {\it slope} in this correlation instead of the 
distance permits the handling of all energies in 
one correlation:
\begin{description}
\item[$\xi$\sc{dis}({\it slope})] is applied to determine Ne(max)
from the observed {\it Ne} at detector level and from {\it slope}
measuring the distance to the shower maximum.
\end{description}
No change of the correlation for different primary energies 
was observed 
(compare table\,\ref{tabene} and the related discussion in the text).
No systematic differences between different primary particles 
are visible.
In the second step the primary energy is determined from Ne(max) 
and E/A.
Two sub steps are necessary here:
first the energy contained in the electromagnetic part of the EAS
has to be derived,
followed by an E/A dependent correction to 
determine the primary energy.
The electromagnetic energy is 
proportional to Ne(max) for a fixed shape of 
the longitudinal shower development. 
Because the shape changes slightly with primary energy E both 
Ne(max) and E are necessary to determine the electromagnetic energy.
During our simulations the total amount of electromagnetic
energy 
in an EAS (defined as the sum of all ${\rm \pi^0}$ energies)
was not recorded so that 
the ratio of electromagnetic energy to the primary energy 
and the ratio of Ne(max) to the electromagnetic energy 
could not be calculated directly for the simulated events.
Hence arbitrary factors may be 
multiplied to the two correction functions below
as long as their product is kept constant.
The functions are:
\begin{description} 
\item[$\xi$\sc{lon}(E)] takes into account the change of 
the longitudinal shower development with E.
\item[$\xi$\sc{em}(E/A)] is used to correct 
from the electromagnetic energy to the total energy E of the 
primary nucleus
and will be discussed a little more detailed. 
\end{description}
The fraction of the primary energy which goes into the 
electromagnetic part of the shower
rises with
increasing energy as the probability for hadrons to 
perform subsequent interaction with the production of 
additional neutral pions increases with the 
hadron energy.
For a nucleus the fraction of the electromagnetic 
energy should depend 
only on E/A.
Following the results 
of \cite{eem} a function was fitted to the 
correlation of 
Ne(max)/E with E/A
using the generated MC events:
\begin{equation}
{\rm \xi\scm{em}(E/A)\,=\, 
\frac{Ne(max)}{E \cdot \xi\scm{lon}(E)}\,=\,
\left[ 1.- \left( \frac{E/A}{33\,GeV} \right)
^{-0.181}\right]\,TeV^{-1}\ .}
\label{eqgab}
\end{equation}
The ratio of this correction
for protons to iron at
300\,TeV amounts to 1.34 decreasing to 1.16 at 5\,PeV.
The difference between proton and iron showers is larger 
than derived by 
extrapolating the results in \cite{eem} 
to the mean atomic number of air because
the fraction of the total energy deposited in
the electromagnetic cascade
is different in air showers compared to 
showers developing in solid state calorimeters:
in air the interaction length for charged pions 
is comparable to their decay length so that
the competition between pion decay and 
secondary interaction with subsequent production of 
neutral pions (feeding the em.\ cascade by their decay 
to two photons)
lowers the 
fraction of energy deposited in the electromagnetic cascade. 
Now the primary energy is calculable by
\begin{equation} 
{\rm E\,=\,\frac{Ne(max)}{\xi\scm{lon} \cdot \xi\scm{em}}
\,=\,\frac{{\it Ne}}{\xi\scm{dis} \cdot \xi\scm{lon} \cdot \xi\scm{em}}\ \ \ .}
\label{eqenemax}
\end{equation}
Due to the energy dependence of the correlation
between {\it slope} and distance to the maximum 
and of Ne(max)/E
(both due to a slightly changing shape of 
the longitudinal shower development)
energy and distance to the shower maximum cannot strictly be determined 
separately but have to be calculated iteratively.
However the energy dependencies are small.
Therefore in the calculation as a start value a parameterization of 
the correlation of distance and {\it slope} neglecting any energy 
dependence is used to derive a first distance value.
With this an energy estimation is calculated and with this energy
a new distance.
This distance in turn gives a new energy value again.
After two iterations usually neither the distance nor the energy
results change further.\\   
The application of the whole procedure to simulated events 
results in systematic uncertainties on the order of 5\%. 
Several contributions to the energy resolution 
for iron and proton showers 
are listed in table\,\ref{tabene}.
\begin{table}[htb]
\begin{center}
\begin{tabular}{|l||r|r||r|r|} \hline
Method & Fe 300\,TeV & 
Fe 5\,PeV & Prot.\ 300\,TeV
& Prot.\ 5\,PeV \\ \hline \hline
Ne(max) & $(6 \pm 1)$\% & $(4 \pm 1)$\% & $(15\pm 1)$\% &
$(9 \pm 1)$\% \\ \hline
Ne at detector, & & & & \\
dist.\ from MC &$ (12 \pm 1)$\%  & $(6\pm 2)$\% & $(17\pm 1)$\% & 
$(7\pm 1)$\% \\ \hline
{\it Ne} from fit,& & & & \\  
dist.\ from MC & $(20\pm 2)$\% & $(7 \pm 2)$\% & $(18\pm 1)$\% & 
$(11\pm 2)$\% \\ 
\hline{\it Ne, slope} from fits,& & & & \\  
E/A from MC & $(22\pm 2)$\% & $(7\pm 2)$\% & $(15\pm 1)$\% & 
$(11\pm 2)$\% \\ \hline
\hline{\it Ne, slope} from fits,& & & & \\  
E/A reconstr. & $(31\pm 3)$\% & $(12 \pm 4)$\% & $(25\pm 2)$\% & 
$(11\pm 2)$\% \\ \hline
\end{tabular}
\caption{{\small Rms values of different quantities
contributing to the energy resolution which can be achieved
with the method using {\it Ne} and {\it slope}.
In the first two rows the identity of the primary 
particle is used from the simulations.
``Ne(max)'', ``Ne at detector'' and ``dist'' are MC quantities,
{\it Ne} and {\it slope} experimental observables. E/A denotes 
the energy per nucleon, which in the last line of the table is 
reconstructed from {\it slope} and the zenith angle.}}
\vspace{-0.5cm}  
\label{tabene} 
\end{center}  
\end{table}
For 300\,TeV iron showers most of the uncertainties 
stem from the NKG fit to the scintillator data with
subsequent fluctuations in {\it Ne} and 
from a modest resolution for E/A (compare Figure\,5).
At 300\,TeV the ratio of reconstructed to generated energy 
exhibits a tail to large values which
originates from the large uncertainties in the determination 
of E/A and the relatively large corrections depending on 
E/A as shown in equation\,\ref{eqgab}.
Already at an energy of 500\,TeV the tail to high energies nearly 
disappears resulting in a rms value of 20\%.
For energies of 1\,PeV and larger the resolution amounts to roughly 10\%.
The much improved energy resolution is 
achieved due to better {\it Ne} and E/A determinations and 
a smaller correction depending on E/A. \\
The energy resolution for proton showers 
improves from 25\% at 300\,TeV to about 10\% at 5\,PeV.
Even a direct measurement of Ne(max) and an unambiguous
identification of its mass would not 
improve the energy resolution very much compared to the reconstruction
using only experimental observables.
\subsection{Energy Reconstruction from 
Cherenkov Light alone} 
\label{ecl}
As in the previous section the energy can be 
reconstructed by replacing 
{\it Ne} by {\it L(20-100)}.
The fraction of the Cherenkov light in the 
interval from 20 to 100\,m depends on the 
distance to the shower maximum (geometry) and on 
E/A, because 
the lateral spread of an EAS decreases with 
decreasing ratio of transverse to longitudinal momentum 
in the interactions.
The following effects were taken into account 
for the energy reconstruction with Cherenkov light only:
\begin{description}
\item[$\zeta$\sc{dis}({\it slope},E):] The fraction of light contained in {\it L(20-100)} 
depends on the distance to the shower maximum.
Due to the differences in the longitudinal
shower development an additional small energy dependence
was also taken into account 
(corresponding to $\xi$\sc{dis} and $\xi$\sc{lon} of the previous section).
\item[$\zeta$\sc{em}(E/A):] For a given distance the fraction in {\it L(20-100)}
depends on E/A also.
As in the previous section ($\xi$\sc{em}(E/A)) this originates
from the energy fraction deposited
in the electromagnetic cascade of an EAS, but 
the lateral extension of the Cherenkov light pool also contributes.
\item[$\zeta$\sc{den}(height):] The threshold for electrons to produce Cherenkov light
varies from 38\,MeV at a height of 300\,g/cm$^2$
to 21\,MeV at the detector level of 793\,g/cm$^2$.
Therefore the amount of Cherenkov light generated in the 
atmosphere depends on the height of the shower maximum.
\end{description}  
The correction depending on energy per nucleon is 
given explicitly below.
\begin{equation}
{\rm \zeta\scm{em}(E/A)}\,=\, 
{\rm
\frac{{\it L(20-100)}\,/\, 1.3 \cdot 10^7}
{E \cdot \zeta\scm{dis} \cdot \zeta\scm{den}}} 
\, =\,
{\rm 
\left[ 1.- \left( \frac{E/A}{178\,GeV} \right)
^{-0.180}\right]\,TeV^{-1}.}
\label{eqlgab}
\end{equation}
This correction is larger than equation\,\ref{eqgab}
because of the 
correction for the Cherenkov light beyond 100\,m distance. 
The remaining systematic uncertainties after all corrections
for different primary particles, energies and zenith angles
are less than 10\%,
a little worse compared to the energy reconstruction with 
{\it Ne}.
The reasons are the  larger E/A dependencies.
The energy resolution ranges from 45\% (35\%) at 300\,TeV
to 8\% (11\%) at 5\,PeV for iron (proton) induced showers. 
\subsection{Comparison of both Energy Reconstructions}
Assuming perfect detectors the light in the interval
from 20 to 100\,m core distance 
can be reconstructed with an error smaller than 1\%.
Clearly this is superior to the measurement of {\it Ne}.
However the obtainable energy resolution at low energies
is limited by 
uncertainties in the distance and E/A reconstruction.
In spite of the    
accurate measurement of the Cherenkov light
the final resolution at 300\,TeV
is even worse than using {\it Ne}
mainly because the correction depending on E/A
is larger.
At high energies fluctuations in the shower 
development concerning its shape and the fraction of energy 
deposited in the electromagnetic cascade limit the 
energy resolution, 
which could only be improved by
an accurate determination of the non electromagnetic component.
It should be noted however that energy resolutions around 10\%
as obtained here for 5\,PeV are already much better 
than needed for most applications.\\
The energy resolutions for both methods
suffer mainly from 
the same uncertainties in distance and E/A determination.
Therefore no significant improvement can be obtained by combining 
both measurements.
However the two methods to determine the primary energy
are not equally sensitive to experimental errors.
For example
the energy reconstructed with Cherenkov light only 
is much less influenced by faulty {\it slope}
measurements than the reconstruction with 
{\it Ne}.
Assuming that the light density at 100\,m
distance from the shower core is measured correctly
while {\it slope} is reconstructed incorrect,
the energy determined with {\it Ne} is incorrect by nearly 
a factor of two if {\it slope} is changed by 20\%
whereas the energy derived from Cherenkov light only
is shifted by less than 10\%.
This different behavior is explained
by the fact, that a too large (small) {\it slope}
leads to too small (large) corrections $\xi$\sc{dis}
and $\zeta$\sc{dis}, but in addition 
too large (small) {\it slopes} give too small (large)
values for {\it L(20-100)} partly compensating 
the effect of the wrong $\zeta$\sc{dis} correction.\\
\section{Determining the Chemical Composition}
\label{chemie}
With the reconstruction of the shower development 
described in the previous section not only the energy 
is inferred independently of the mass of the 
primary nucleus but also the nucleon number of the hadron hitting 
the atmosphere can be determined coarsely by only 
measuring the electromagnetic part of the EAS.
In the following sections first the 
energy per nucleon derived from the longitudinal
shower development 
and then the properties of the lateral shower extensions
will be analyzed for their sensitivity on the nucleon number of 
the primary particle.
The third section combines all information concerning the 
chemical composition deduced here from the four observables 
{\it Ne, age, slope} and {\it L(20-100)}
for 300\,TeV showers as an example.
\subsection{The Nucleon Number from the \protect\newline
longitudinal Shower Development}
With procedures to determine the primary energy 
and an estimation of the energy per nucleon E/A 
from the position of the shower maximum as described 
in section\,\ref{detea}
it is straightforward to calculate the nucleon number:
\begin{equation}
{\rm log_{10}(A) \, \equiv \, 
log_{10}\left(\frac{energy}{energy/nucleon}\right)
.}
\label{loga}
\end{equation}
Figure\,\ref{massrec} displays the reconstructed 
log$_{10}$(A) values for proton and iron showers 
of 0.3 and 5\,PeV.
The other primaries were omitted in order to keep
a clear picture.
     \begin{figure}[htb]
     \begin{center}
     \vspace{-1.3cm}
     \epsfxsize=10.55cm
     \epsfysize=5cm
     \mbox {\epsffile{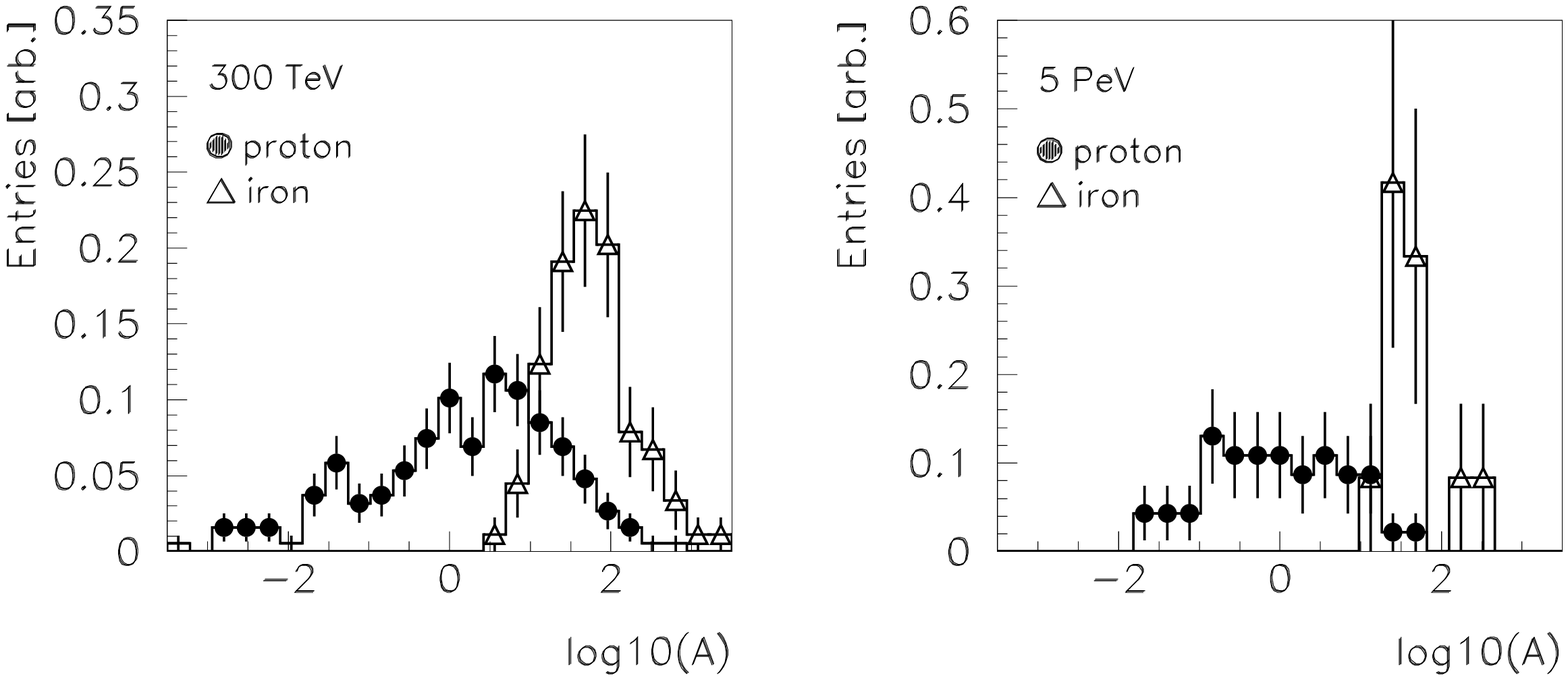}}
     \vspace{-0.5cm}
     \caption{{\small The reconstructed nucleon number 
for proton and iron primaries of 300\,TeV (left)
and 5\,PeV (right).}}
      \vspace{-0.5cm}
     \label{massrec}
     \end{center}
   \end{figure} 
Further results are summarized in table\,\ref{tabmass}.
The reconstructed mean values 
correspond within statistical errors to the 
expectation values. 
For the energy reconstruction the method with 
{\it Ne} was used, but the same results are obtained 
with the second method of determining
the primary energy.\\ 
Light and heavy primaries
\begin{table}[htb]
\begin{center}
\begin{tabular}{|lc||c|c||c|c|} \cline{2-6}
\multicolumn{1}{c}{\ } & 
\multicolumn{5}{||c|}{\rule{0cm}{0.5cm} 
\raisebox{0.1cm}{log$_{10}$(A)}} \\ \hline
Primary & \multicolumn{1}{||c||}{Mean} & RMS 300\,TeV & RMS MC & 
RMS 5\,PeV & RMS MC \\ \hline \hline
Proton & \multicolumn{1}{||c||}{0.00} & $1.19 \pm 0.09 $  & $1.00 \pm 0.07 $ 
& $0.82 \pm 0.13 $ & $0.77 \pm 0.12 $ \\
Helium & \multicolumn{1}{||c||}{0.60} & $0.83 \pm 0.08 $ & $0.70 \pm 0.07 $ 
& $0.42 \pm 0.12 $ & $0.30 \pm 0.09 $ \\
Oxygen & \multicolumn{1}{||c||}{1.20} & $0.63 \pm 0.08 $ & $0.52 \pm 0.07 $ 
& $0.26 \pm 0.08 $ & $0.22 \pm 0.06 $ \\
Iron   & \multicolumn{1}{||c||}{1.75} & $0.53 \pm 0.06 $ & $0.35 \pm 0.04 $ 
& $0.38 \pm 0.11 $ & $0.29 \pm 0.09 $ \\ \hline 
\end{tabular}
\caption{{\small The mean and rms values of the 
distributions of the reconstructed log$_{10}$(A) values.
``MC'' symbolizes the result obtained by using the generated
MC energy and the depth of the shower maximum directly from MC.
The numbers given in the ``MC'' columns therefore show the 
contributions from fluctuations in the longitudinal
shower development only.
Differences to the fits shown in Figure\,5 originate
from the summation over all zenith angles in this table.}}  
\vspace{-0.5cm}
\label{tabmass} 
\end{center}  
\end{table}  
can be distinguished by their 
different mean values and by their different spreads.
The spread of the ${\rm log_{10}(A)}$ 
distributions is dominated by the statistical 
fluctuations of the depth of the shower maximum 
with subsequent uncertainties in the E/A determination
(see Figures\,\ref{detenuc} and 5).
Even a perfect energy determination would hardly improve 
the separation of different primary particles.
\subsection{Composition Analysis from the  \protect\newline
lateral Shower Development}
\label{complat}
In section\,\ref{showerlat} differences concerning 
observable lateral extensions of EAS which depend on E/A
have been touched briefly.
{\it Age} can be used to estimate E/A if the 
energy of the primary particle and the distance to the shower maximum is known.
Figure\,\ref{compae} (left) compares {\it age} for different 
primaries of 300\,TeV energy. 
To use this discrimination the expectation value of 
{\it age } for proton induced showers was parameterized:
\begin{equation}
{\rm { age}(p)\, =\, 1.42 - 0.10 \cdot log_{10}(E/TeV)
+ 18.0 \cdot {\it slope}\ .}
\label{agep}
\end{equation}
For each reconstructed shower the actual {\it age} is then 
compared to the expectation for primary protons. At 300\,TeV
iron primaries show a mean value {\it age/age(p)} of 
1.20 with a rms of 0.17 while the mean for protons lies at 1.00
as expected with a rms of 0.15.\\
EAS of different nuclei can also be distinguished by their 
lateral shower developments as measurable 
by comparing the number of Cherenkov photons within and beyond 
100\,m core distance. 
Unfortunately it is very difficult to measure the low density Cherenkov light
up to a few hundred 
meters distance from the shower core with great precision.
However using the energy reconstruction methods 
developed in this paper an indirect measurement of the light 
beyond 100\,m is possible:
if the energy is reconstructed only with Cherenkov light  
an E/A dependent correction (eq.\,\ref{eqlgab}) has to 
be applied to take into account the changing fraction of 
Cherenkov light measurable below 100\,m and the fraction of 
the primary energy deposited in the electromagnetic cascade.
Only the latter point has to be corrected if the energy 
reconstruction is done with the help of {\it Ne} (eq.\,\ref{eqgab}).
Therefore omitting all E/A corrections in both energy
reconstruction methods (resulting in E$^*$(Cl) and E$^*$({\it Ne}))
and then comparing E$^*$(Cl)/E$^*$({\it Ne})
provides an indirect estimation of the amount of 
Cherenkov light at large distances.
This is equivalent to compare the number of Cherenkov photons 
between 20 and 100\,m core distance to {\it Ne}
taking into account the distance to the shower maximum and density 
effects for the production of Cherenkov light.
In Figure\,\ref{compae} (right) the energy ratios are plotted.
A clear separation 
is visible for 300\,TeV showers. 
     \begin{figure}[htb]
     \begin{center}
     \vspace{-1.3cm}
     \epsfxsize=10.55cm
     \epsfysize=5cm
     \mbox {\epsffile{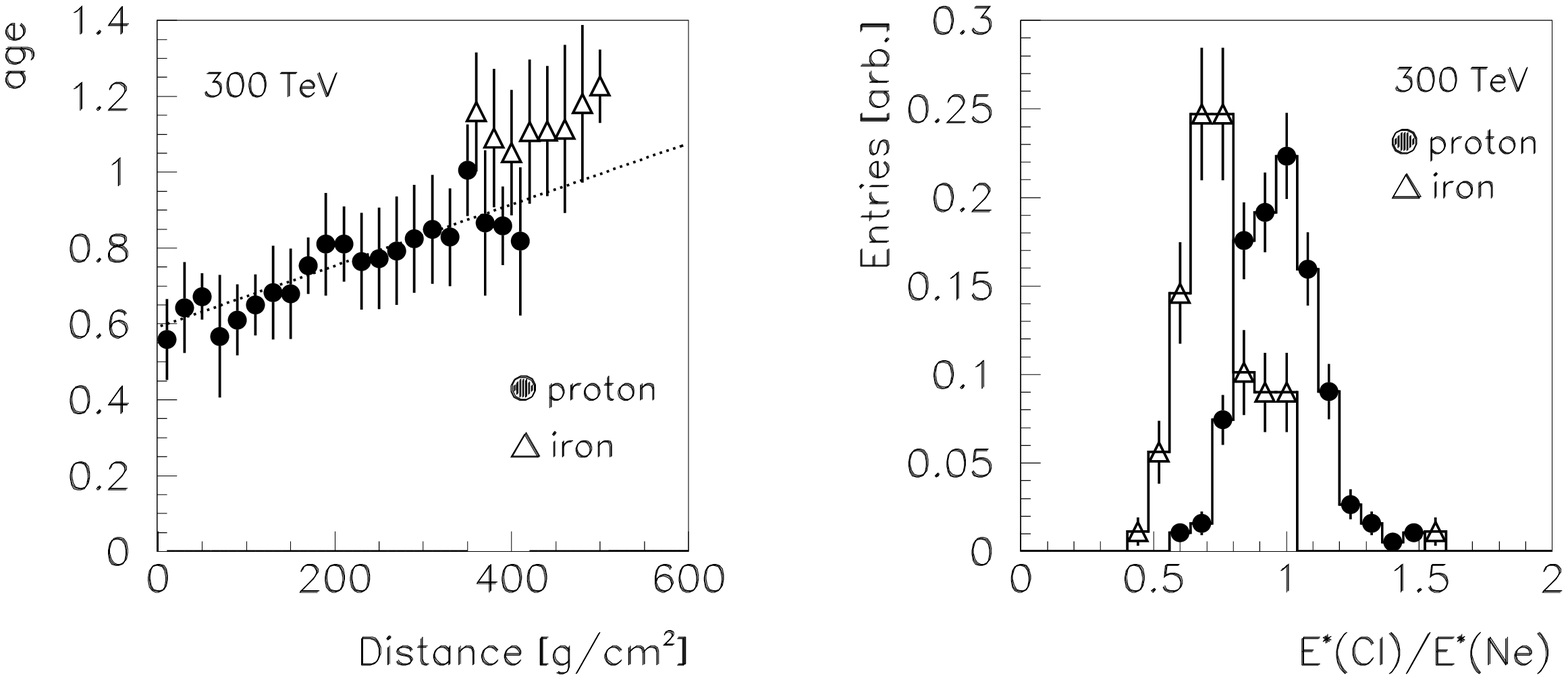}}
     \vspace{-0.5cm}
     \caption{{\small The {\it age} values  
as a function of the distance to the 
shower maximum (left), and
the ratio of the energies reconstructed only with
Cherenkov light and by using {\it Ne} and {\it slope}, where
all corrections depending on E/A were omitted (right).
Proton and iron showers of 300\,TeV are displayed.}}
     \vspace{-0.5cm}
     \label{compae}
     \end{center}
   \end{figure}
An analysis of the chemical composition can 
profit from measuring the lateral extent of 
the electromagnetic cascade of an EAS at energies below 
approximately 1\,PeV.
At higher energies the lateral extensions
no longer depend on the primary mass
in a measurable way.
With the observables used in this paper
the chemical composition around the ``knee''
can only be derived from the longitudinal shower development.
\subsection{Chemical Composition from a combined 
Analysis of the longitudinal and lateral 
Shower Development }
The sample of 300\,TeV showers was used to compare the 
sensitivity of the different parameters discussed in the last 
two sections on the mass of the primary nucleus. 
If cuts in different quantities are applied 
so that 90\% of the iron showers are selected the following fractions 
of proton shower remain:
\begin{itemize}
\item Cut in log$_{10}$(A): 20\%,
\hspace*{0.5cm} $\bullet$ \ cut in {\it age}/age(p) or 
E$^*$(Cl)/E$^*$({\it Ne}): 40\%.
\end{itemize} 
The most sensitive 
parameter is derived from the longitudinal 
shower development, but also a discrimination between 
heavy and light nuclei can be derived from the lateral shower 
extensions.
To combine the information from the longitudinal shower 
development and the two comparisons referring to lateral
extension of the EAS the probability densities for observing
a specific log$_{10}$(A), E$^*$(Cl)/E$^*$({\it Ne}) 
or {\it age/age(p)} value were 
parameterized for primary p, ${\rm \alpha}$, O and Fe nuclei of 
300\,TeV from the MC library.
Following equation\,\ref{eqallco} 
(similar for other primaries than iron) a combined probability is 
calculated:\\
${\rm With \ \rho_{Fe}\, =\, prob({\it log_{10}(A)},Fe) \cdot 
prob({\it E^*(Cl)/E^*(Ne)},Fe) \cdot prob({\it age/age(p)},Fe)}$ 
\begin{equation}
{\rm prob(Fe)\,=\,  
\frac{\rho_{Fe}}{\rho_{p}+\rho_{\alpha}+\rho_{O}+\rho_{Fe}}} 
\label{eqallco}
\end{equation}
Table\,\ref{taballco} lists the fractions for nuclei of 
different masses which are obtained by selecting 90\% or 50\% of all
proton or iron showers.
\begin{table}[htb]
\begin{center}
\begin{tabular}{|l||r|r||r|r|} \hline
Primary & \multicolumn{1}{|c||}{Sel.\ 90\%\,p} 
& \multicolumn{1}{|c|}{Sel.\ 50\%\,p} 
& \multicolumn{1}{||c|}{Sel.\ 90\%\,Fe} 
& \multicolumn{1}{|c|}{Sel.\ 50\%\,Fe} 
\\ \hline \hline
Proton & 90\% \hspace*{0.8cm} & 50\% \hspace*{0.8cm}
& 8\% \hspace*{0.8cm}& $<$1\% \hspace*{0.8cm}  \\
Helium & 80\% \hspace*{0.8cm}& 17\% \hspace*{0.8cm}
& 12\% \hspace*{0.8cm}& 1.5\% \hspace*{0.8cm}\\
Oxygen & 43\% \hspace*{0.8cm}& 3\% \hspace*{0.8cm}
& 60\% \hspace*{0.8cm}& 24\% \hspace*{0.8cm}\\
Iron   & 5\% \hspace*{0.8cm}& $<$1\% \hspace*{0.8cm}
& 90\% \hspace*{0.8cm}& 50\% \hspace*{0.8cm}\\ \hline
\end{tabular}
\caption{{\small The remaining fraction of primaries with energies of 300\,TeV
after selecting 
90\% or 50\% of the proton (iron) showers with cuts in prob(p) 
or prob(Fe).}}
\vspace{-0.5cm}
\label{taballco} 
\end{center}  
\end{table}  
Clearly an analysis of the chemical composition 
improves if measurements of the 
longitudinal and lateral shower developments are combined.
Light and heavy particles can be separated rather well. However 
with the four observables used here primary nuclei with  
masses similar to oxygen can be separated only in a statistical sense
but not on an event by event basis.
It seems to be difficult to distinguish between 
primary protons and ${\rm \alpha}$ particles. \\
\subsection{Systematics}
\label{wo}
Studies of systematic effects related to the CORSIKA code
(stepwidth of the EGS part), atmospheric transmission,
the fragmentation of the primary nucleus and the influence 
of different models to simulate the high energy interactions  
of the CR will be described in a forthcoming publication.
In general the observables analyzed here show up to be much less 
model dependent than hadronic shower properties or muon
distributions, mainly because the development of a
shower behind its maximum determines the Cherenkov light 
and particle measurement, as considered throughout this paper.
\section{Summary and Conclusions}
In this paper methods were presented to determine energy and mass
of charged cosmic rays from ground based observations
of the electromagnetic cascade of air showers.
From the slope of the lateral Cherenkov light density in the range 
of 20 to 100\,m core distance the position of the shower maximum 
can be inferred without knowledge of the nucleon number of the 
primary particle.
This leads to an unbiased determination of the energy per nucleon
and, combined with the shower size at 
detector level or the number of 
registered Cherenkov photons,
to a measurement of the primary energy.
Thus a measurement of the energy spectrum 
and a coarse determination of the chemical composition
are possible
without any a priori hypotheses. \\
With the observables considered in the present paper
the energy resolution for primary nuclei is limited to 
approximately 30\% at 300\,TeV improving to 10\% at 5\,PeV
due to natural fluctuations in the shower development.
Further improvements of these results are only possible
if accurate measurements of the non electromagnetic components 
of EAS are added.\\
At energies below 1\,PeV, where results from EAS measurements 
can be compared to direct data from balloon flights, 
the sensitivity of the analysis of air showers by observing 
Cherenkov light and particles at detector level
can be substantially improved by combining 
the results related to the longitudinal shower 
development with parameters derived from the lateral extension.
This allows detailed tests of the described method to determine 
the chemical composition and the energy spectrum of cosmic rays. \\
One main characteristic of deriving energy and mass of primary nuclei
from observations of the electromagnetic component of extensive air 
showers with the observables used here is the fact, 
that it are mainly the longitudinal shower 
development behind the shower maximum,
the number of particles at the maximum
and the penetration depth of the shower until it reaches the maximum,
which determine the results.
While the first two items do not vary much for 
different models describing the development of air showers
the last item is more model dependent.
In order to achieve results being as model independent as possible
it is very desirable to combine the method described in this paper 
with complementary measurements.
Analyses of the early stage of the shower development, of the 
hadronic component of EAS or detailed studies of the shower core
may be considered for this purpose. 
\vspace{0.5cm} \\
{\bf ACKNOWLEDGMENTS}
\vspace{0.2cm}\\
The author would like to thank the HEGRA members for 
their collaboration.
Especially I am very grateful to V.\,Haustein, 
who performed most of the MC simulations and determined the
chemical composition of CR with a different technique,
to G.\,Heinzelmann for many detailed, constructive
proposals and improvements as well as for his general support,
and to R.\,Plaga, who pioneered the analysis of charged CR 
in the HEGRA collaboration, for motivations and detailed discussions.  
Special thanks to the authors of CORSIKA for
supplying us with the simulation program and their support.
I thank Gerald Lopez for his careful reading of the text and 
for providing valuable suggestions.
This work was supported by the BMBF (Germany) under contract number
05\,2HH\,264.

}
\end{document}